\documentclass[12pt]{iopart}

\usepackage{graphics}
\usepackage{graphicx}
\usepackage{enumerate}

\begin{document}

\title{Real-time capable first principle based modelling of tokamak turbulent transport}

\author{J.~Citrin$^{1,2}$, S.~Breton$^2$, F.~Felici$^3$, F.~Imbeaux$^2$, T.~Aniel$^2$, J.F.~Artaud$^2$, B.~Baiocchi$^4$, C.~Bourdelle$^2$, Y.~Camenen$^5$, J.~Garcia$^2$}

\address{$^1$FOM Institute DIFFER -- Dutch Institute for Fundamental Energy Research, P.O. box 6336, 5600 HH Eindhoven, The Netherlands}
\address{$^2$CEA, IRFM, F-13108 Saint Paul Lez Durance, France}
\address{$^3$Eindhoven University of Technology, Department of Mechanical Engineering, Control Systems Technology Group, P.O. Box 513, 5600MB, Eindhoven, The Netherlands}
\address{$^4$Istituto di Fisica del Plasma ``P. Caldirola'', Associazione Euratom-ENEA-CNR, Milano, Italy}
\address{$^5$Aix-Marseille Universit´e, CNRS, PIIM UMR 7345, 13397 Marseille, France}
\ead{J.Citrin@differ.nl}

\begin{abstract}
A real-time capable core turbulence tokamak transport model is developed. This model is constructed from the regularized nonlinear regression of quasilinear gyrokinetic transport code output. The regression is performed with a multilayer perceptron neural network. The transport code input for the neural network training set consists of five dimensions, and is limited to adiabatic electrons. The neural network model successfully reproduces transport fluxes predicted by the original quasilinear model, while gaining five orders of magnitude in computation time. The model is implemented in a real-time capable tokamak simulator, and simulates a 300~s ITER discharge in 10~s. This proof-of-principle for regression based transport models anticipates a significant widening of input space dimensionality and physics realism for future training sets. This aims to provide unprecedented computational speed coupled with first-principle based physics for real-time control and integrated modelling applications.
\end{abstract}


\textit{Introduction}.-- Particle, momentum, and heat transport in the tokamak core is dominated by turbulence driven by plasma microinstabilities~\cite{hort99,ITER2}. An accurate predictive model for turbulent transport fluxes is thus vital for the interpretation and optimization of present-day experiments, and extrapolation to and control of future machines. 

Direct numerical simulation with massively parallel nonlinear gyrokinetic codes has provided tremendous insight to the underlying transport physics and success in reproducing experimental fluxes in many regimes. However, the computational cost -- typically $10^5$ CPU hours for a local flux calculation at a single radial point -- precludes the routine use of such codes for integrated tokamak transport simulations which demand $\sim10^3$ flux computations per 1~s of plasma evolution on JET scale devices. 

Reduced turbulent transport models have been constructed to increase tractability. They are based on the quasilinear approximation, which is proven to be largely valid in the core of tokamak plasmas~\cite{lin07,dann05,casa09}. These rely on nonlinear simulations for validating their ansatzes and normalizing factors. These models have proven successful in reproducing experimental profiles in many cases. Examples are TGLF~\cite{stae07} and QuaLiKiz~\cite{bour07}. A $\sim$6 orders of magnitude speedup is gained in quasilinear calculations compared to nonlinear simulations. However, while extremely useful, the tractability of such models is still marginal for convenient large-scale scenario development over discharge timescales. For example, 1~s of JET plasma evolution can take up to 10 hours with 10 processors, depending on the integrated modelling platform used. This speed is also insufficient for applications such as trajectory optimization, and simulations for developing real-time controllers. Furthermore, any increase in physics fidelity in the models often results in a trade off with further decrease in tractability.

This Letter illustrates an approach to overcome these challenges. The central point is to relegate the expensive flux calculations to a stage precedent to its use in a transport simulation. Instead, analytical formulae are to be used in the simulation, based on a neural network (NN) nonlinear regression of quasilinear fluxes previously compiled in a database. The advantage is twofold: 1) the numerical resolution of analytical formulae is orders of magnitude faster than original flux calculation; 2) the computation time required for compiling the database is independent from the computation time spent during the tokamak simulation itself, hence the training set for NN regression can include results from more complete codes than used in contemporary integrated transport modelling.

Neural networks have found multiple applications in tokamak research, including: nonlinear regression for energy confinement scaling~\cite{alle92}; neoclassical transport~\cite{waka07}; rapid determination of equilibria~\cite{list91}, electron temperature profiles~\cite{clay13}, and charge exchange spectra~\cite{sven99}; classification of disruption~\cite{wrob97,cann04,vega14} and L-H transition onsets~\cite{gaud14}. Most related to this work is a regression of DIII-D heat fluxes from experimental power balance databases~\cite{mene14}. 

\textit{Quasilinear transport model and training set}.-- The QuaLiKiz quasilinear gyrokinetic transport model~\cite{bour07,casa09,citr12,cott14,bourhdr} was employed in this work. QuaLiKiz solves a linear gyrokinetic dispersion relation for calculating wavenumber spectra of instability growth rates and frequencies. Then, integrating over the spectra, the transport fluxes are calculated via quasilinear flux integrals and nonlinear saturation rules. The bulk of the computational time is spent in the first stage, the dispersion relation solver. QuaLiKiz has been coupled to the CRONOS~\cite{arta10} integrated modelling suite, and has successfully reproduced temperature and density profiles of JET and Tore-Supra discharges~\cite{casaphd,baio15}. Following recent upgrades~\cite{citr14}, the computational time for the QuaLiKiz eigenvalue solver at a single wavenumber in QuaLiKiz is on the order of $\sim$1~s.

A database of QuaLiKiz solutions was constructed, in the ion temperature gradient (ITG) instability regime. This instability is often the primary driver of tokamak microturbulence. The code was run with adiabatic electrons for simplicity, which also decreases the computational time to $\sim$300~ms. The database covers four input parameters known to have significant impact on ITG transport fluxes in this regime: the driving normalized logarithmic ion temperature gradient $R/L_{Ti}$, the ion to electron temperature ratio $T_i/T_e$, the safety-factor $q$, and the magnetic shear $\hat{s}\equiv\frac{r}{q}\frac{dq}{dr}$. In addition, the input normalized wavenumber $k_\theta\rho_s$was scanned, constricted to above ion-Larmor-radius scales, where $\rho_s\equiv\sqrt{T_em_i}/(Z_iq_eB)$. The following parameters were maintained fixed: the normalized logarithmic density gradient $R/L_n=3$, normalized radial location $r/a=0.5$. No Shafranov shift was assumed in the geometry. The database consists of a dense grid of points summarized in table~\ref{tab:summary1}, from which the training sets for the neural network were sifted. The QuaLiKiz outputs we investigate are: growth rates and frequencies, which correspond to 5D input space; ion heat flux, which corresponds to 4D input space due to integration over wavenumbers. The database includes cases corresponding to unstable modes, and cases where no instabilities were found by QuaLiKiz, and the outputs are set to 0. 

A regression of the ion heat flux has immediate application for transport modelling. However, a regression of the more primitive linear output has its own specific applications. For example, since the dispersion relation solver is the slowest part of the code, a fast reproduction of growth rates, frequencies and eigenfunctions would allow rapid tests of various saturation rule formulations throughout parameter space. These saturation rules typically evolve following continuous comparisons with nonlinear simulations and experiments. In this sense, a database consisting of the complete outputs of linear codes does not become obsolete, while a quasilinear flux database can. 
\begin{table}
\centering
\caption{\footnotesize Summary of input parameters for the QuaLiKiz adiabatic electron ITG database employed in this work}
\tabcolsep=0.09cm
\scalebox{0.84}{\begin{tabular}{c c c c}
\label{tab:summary1}
Parameter  & Min value & Max value & No. of points \\
\hline
$R/L_{Ti}$& 2 & 12 & 30 \\
$T_i/T_e$& 0.3 & 3 & 20 \\
$q$& 1 & 5 & 20 \\
$\hat{s}$& 0.1 & 3 & 20 \\
$k_\theta\rho_s$& 0.05 & 0.8 & 16 \\
\hline
\multicolumn{3}{c}{Total no. of points} & 3 840 000\\
\hline
\end{tabular}}
\end{table}

\textit{Neural networks}.--  The goal is to find analytical formulae which robustly reproduce the various QuaLiKiz outputs. To this end, a multilayer perceptron neural network is used, which is a nonlinear function with tunable variables (weights and biases), with the property of universal approximation~\cite{bish95book,hayk98book}. For an overview, with an emphasis on applications for fusion, see Ref.\cite{bish94}. Linear combinations of the inputs and biases are propagated through a series of nonlinear transfer function vectors (named `hidden layers'), until eventually linearly combined to an output layer. With two hidden layers and a single output value (as used in this work), this is represented as:
\begin{equation}
\label{eq:NN}
y = b_3  + \sum_i^N{w^2_i}g\left(b^2_i + \sum_j^Mw^1_{ij}g\left(b^1_j+\sum_k^{I}w^{in}_{jk}x_k\right)\right)
\end{equation}
Where $y$ is the output `neuron' containing the output value (i.e. growth rate, frequency, or ion heat flux), $x_k$ the vector of input values, $b^n$ the bias vectors, $w^{in}$ the M$\times$I weight matrix connecting the input vector to the 1st hidden layer, $w^{1}$ the N$\times$M weight matrix connecting the two hidden layers, and $w^{2}$ the weight vector connecting the 2nd hidden layer to the output neuron. $g$ is the nonlinear transfer function, defined as a sigmoid in this work:
\begin{equation}
\label{eq:sig}
g(x)=\frac{2}{1+e^{-2x}}-1
\end{equation}
Following a series of optimization tests, two hidden layers, as shown in equation~\ref{eq:NN}, were employed here. The hidden layer sizes M and N were set to 40. The input layer size, I, is 4 for ion heat fluxes, and 5 for growth rates and frequencies. 

The key stage is the determination of the optimized values of the weights and biases. This is done by minimizing a cost function consisting of the average squared error between the network output and known target output. This set of target output, known as the `training set', is a subset of the QuaLiKiz output values from the database. The BFGS algorithm~\cite{denn96book}, implementing a quasi-Newton method, was used for the weight and bias optimization. All NN training in this work was carried out with the MATLAB neural network toolbox~\cite{matlab}. Following training, the network output then emulates the original model within the database input parameter envelope. This is validated by comparison to validation sets sifted from the database, which are different from the training set.

To avoid overfitting the data, regularization techniques were used in the regression. This corresponds to adding a penalty term in the cost function related to the sum of squares of the network weights and biases, leading to smoother output. The use of regularization ensures that the NN response is smooth (e.g. without strong oscillations) in sparse regions of training set parameter space or when extrapolating beyond the training set envelope.

The analytic form of the nonlinear regression function allows for the calculation of analytical gradients of the outputs with respect to the inputs. This is vital for the efficient solution of fast implicit schemes in real-time capable core transport simulators such as RAPTOR~\cite{feli12}. The regularization also ensures smooth gradients throughout parameter space, important for the stability of such implicit schemes.



\textit{Regression results}.-- We focus on the ion heat flux NN regression, due to its direct relevance for transport modelling applications. Successful regressions of the growth rates and frequencies were also obtained but for brevity not discussed here. 

To capture the instability thresholds with high fidelity, the regression was only carried out for a training set corresponding to unstable modes. The NN output for the stable regions in the validation set was then negative, since the regularized network tends to smoothly extrapolate the trends observed towards the training set envelope. For the final heat flux output, these negative values were then set to zero to represent stability. This scheme avoids having the regularized regression network attempt to directly fit the discontinuous gradients at the instability thresholds, which would be performed poorly due to to the regularization constraint. This is an important point since tokamak transport often tends to be maintained near the critical temperature gradient thresholds, especially in high temperature regimes.

\begin{figure}[htbp]
	\centering
		\includegraphics[scale=0.8]{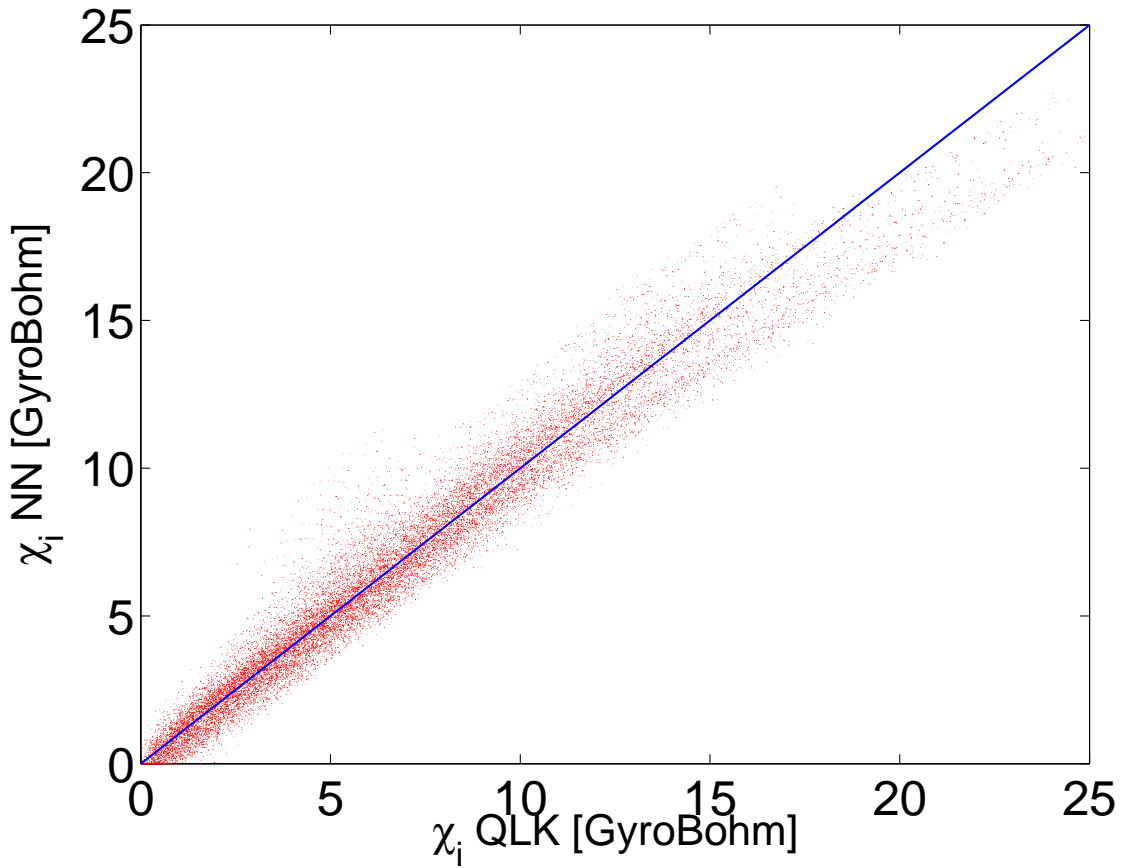}
		\caption{Comparison between normalized ion heat fluxes obtained directly from QuaLiKiz (x-axis) and those from its NN regression (y-axis)}
	\label{fig:figure1}
\end{figure}

The network was trained with a training set of 35000 points chosen randomly from the set of unstable modes in the database. A comparison between the regression NN and QuaLiKiz outputs for a validation set of 10000 unstable cases (different from the training set) is shown in figure~\ref{fig:figure1}. The regression network has an RMS error of 0.77 in gyroBohm units ($\chi_{iGB}=\frac{T_i^{3/2}m_i^{1/2}}{(Z_iq_eB)^2r}$) when compared to the validation set. This RMS error is similar for the training set itself and is primarily due to the regularization constraint. The impact of this error on the simulated profiles is minor. This is due to stiffness, defined here as the local gradient of $\chi_i$ with respect to the driving $R/L_{Ti}$. To quantify this, a comparison was made between the $R/L_{Ti}$ values predicted by the NN and QuaLiKiz to balance a representative $\chi_{iGB}=1$. This was done for all values of $q$, $\hat{s}$, and $T_i/T_e$  in the database. The RMS error was ${\Delta}\left(R/L_{Ti}\right)=0.29$, which corresponded to an average relative error of only 4.2\%.


\begin{figure}[htbp]
	\centering
		\includegraphics[scale=0.8]{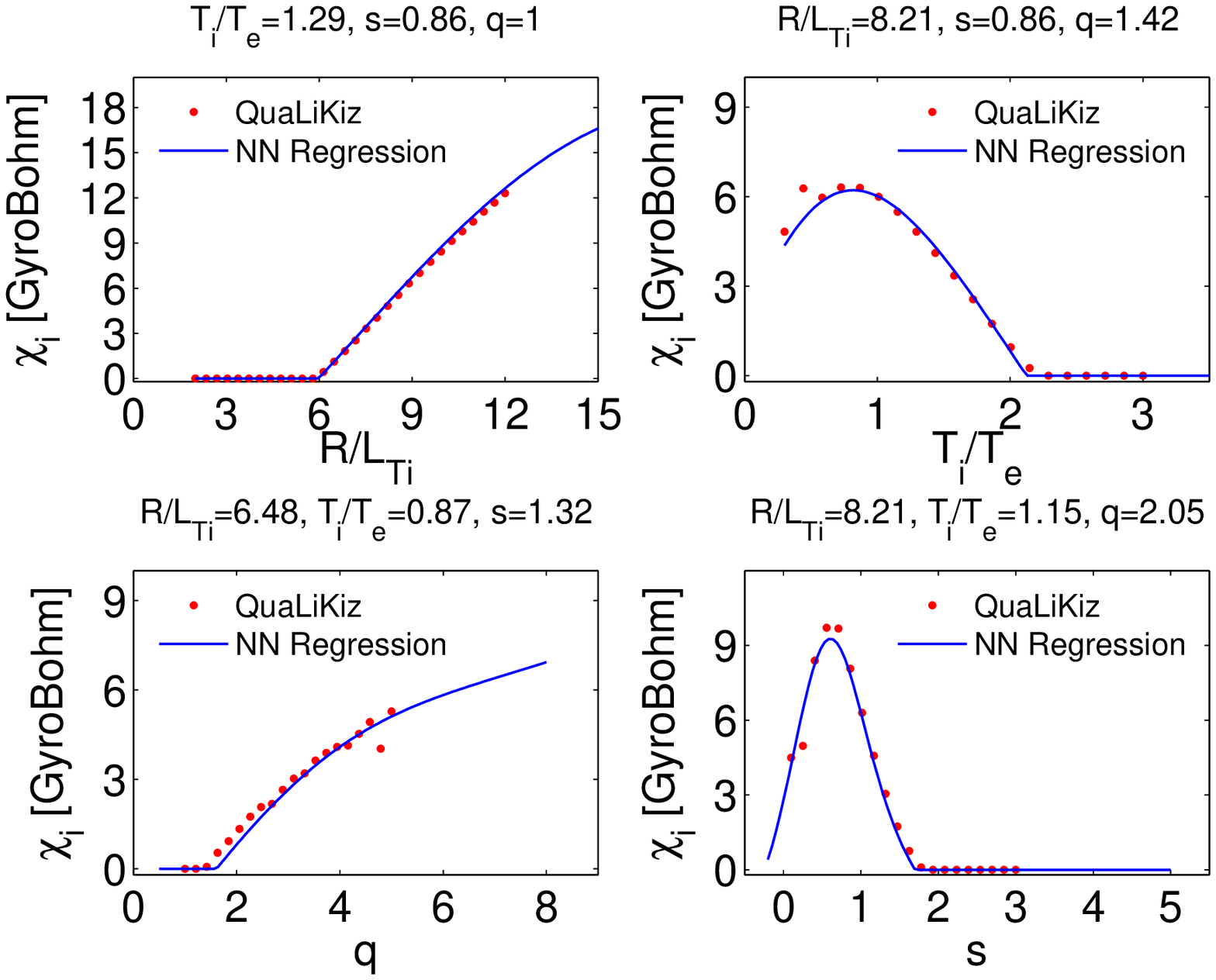}
		\caption{Comparison of NN parameter scans (blue solid lines) vs the original QuaLiKiz ion heat flux calculations (red dots). The scans are in $R/L_{Ti}$ (top left panel), $T_i/T_e$ (top right panel), $q$ (bottom left panel) and $\hat{s}$ (bottom right panel)}
	\label{fig:figure2}
\end{figure}

The typical quality of the fits can be seen in figure~\ref{fig:figure2}, displaying scans of the 4 separate input parameters while the others remained fixed. Negative outputs of the NN network are set to zero. Note the resulting excellent fit of the instability thresholds. In addition, extrapolating the NN scans beyond the range of the training set maintains the trend observed in the data, due to the regularization. This is very encouraging with regard to extension of this approach to more sparse datasets in higher dimensions. However, we do not intend to routinely use NN models in poorly represented regions of parameter space, as the quality of extrapolation cannot be determined a priori.  Rather, the training sets should be continuously expanded to cover such encountered sparse or empty regions, and the NN then periodically retrained. Nevertheless, the smoothness of the regularized NN response when extrapolating ensures its robustness and stability during practical use as a transport model, including during phases when such sparse regions are encountered.”

Each NN output is calculated on a sub 10~${\mu}s$ timescale in MATLAB on a Intel(R) Xeon(R) E5450 CPU @ 3.00GHz. This is a 5 order of magnitude speedup in comparison to the original QuaLiKiz calculations.

\textit{Application in transport codes}.--A transport model based on the trained neural network was constructed, and implemented both in the CRONOS~\cite{arta10} and RAPTOR~\cite{feli12} integrated modelling codes. 

In CRONOS, the validity of the NN transport model was assured by a successful comparison with a JET baseline H-mode shot 73342 with ion and electron heat transport previously simulated~\cite{baio15} with the full QuaLiKiz model. For brevity we will not focus further on this case. Rather, we focus on the real-time simulation capabilities offered by coupling the NN model to RAPTOR. 

Presently, RAPTOR only models electron heat transport. The NN model output was thus modified to roughly approximate ITG regime electron heat transport with kinetic electrons. This was done by assuming that heat fluxes in ITG kinetic electron cases are higher by factor 3 compared with adiabatic electron cases, and furthermore assuming an ion to electron heat flux ratio of $q_i/q_e=3$. These approximations are based on typical nonlinear and quasilinear observations in the ITG regime~\cite{kins05,casa09}. 

\begin{figure}[htbp]
	\centering
		\includegraphics[width=0.6\textwidth]{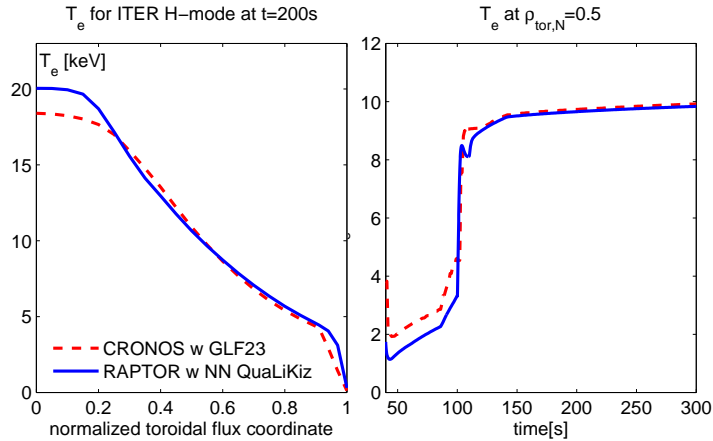}
		\caption{Comparison between the $T_e$ predictions for an ITER hybrid discharge carried out with CRONOS/GLF23~\cite{citr10} (red curve) and a RAPTOR simulation using the QuaLiKiz NN transport model (blue curve). A typical H-mode profile (left panel) and time dependence at mid-radius (right panel) are shown. The LH transition was set at 100~s.}
	\label{fig:figure3}
\end{figure}

In figure \ref{fig:figure3}, we compare a RAPTOR simulation of an ITER hybrid scenario, using the QuaLiKiz NN model for electron heat transport, with a simulation of the same case originally carried out~\cite{citr10} using CRONOS and the GLF23~\cite{walt97} transport model. Using GLF23 allows to compare over ITER-scale discharge times of $>$100s, which is less tractable using the original QuaLiKiz model. For heat transport in a pure ITG regime, GLF23 and QuaLiKiz predictions are expected to be similar, as illustrated in specific single-time-slice comparisons~\cite{baio15}.

The RAPTOR simulation uses all the same actuator (source) inputs and density evolution as the CRONOS simulation. Ion temperatures were held fixed at $T_i/T_e\sim0.8$ in L-mode and $T_i/T_e\sim0.9$ in H-mode. The NN model was operational within a normalized toroidal flux coordinate ($\rho$) range of 0.25 to 0.95. For $\rho>0.95$, $\chi_e$ was feedback controlled to maintain a prescribed edge pedestal temperature of 4~keV. For $\rho<0.25$, a constant $\chi_e$ was assumed to maintain a reasonable level of transport, since GLF23 and QuaLiKiz both predicted stability within that region. A RAPTOR simulation of an entire 300~s ITER discharge took 10~s on a single CPU, corresponding to 30x faster than real-time. This combination of simulation speed and first-principle modelling is unprecedented. With CRONOS/GLF23, the same simulation took 24 hours. 



\textit{Conclusions and outlook}.-- A neural network fit to a restricted subspace of quasilinear gyrokinetic transport model calculations, relevant in the ITG regime, was carried out and applied as a transport model for integrated modelling. While the quasilinear model, QuaLiKiz, is 6 orders of magnitude faster than nonlinear simulations, the NN regression leads to a further 5 order of magnitude speedup. This model is thus real-time capable while still being based on first-principles, which is unprecedented. This model has been coupled to the CRONOS integrating modelling suite, and validated against a full QuaLiKiz simulation in the ITG regime. The model is also coupled to the real-time capable RAPTOR tokamak simulation code, and can model a 300s ITER discharge within 10s, with good agreement with previous modelling using CRONOS and the GLF23 transport model. 

This opens up many new possibilities for real-time controller design and validation, scenario preparation and optimization, and real-time discharge supervision. Such models can be used to design controllers for the plasma profiles using model-based controller design methods (e.g. \cite{malj15} or \cite{more13}). The transport model can be used in closed-loop simulations to validate the designed controllers. Recent work on plasma ramp-up trajectory optimization \cite{vDon14} was carried out with an ad-hoc transport model, and can now be improved using this first-principle-based transport model. Also, this transport model can be used in real-time simulations to verify the measured plasma evolution and warn a supervisory control system of any unexpected deviations during the discharge \cite{feli12b}. Specifically for ITER, the faster-than-real-time opens up the possibility of (on-line) real-time optimization of the discharge evolution in response to such unexpected events.

While applications in the ITG transport regime are already feasible with this model, there remains much scope for expanding the number of input dimensions in the databases used for the fits, as well as employing slower yet more complete linear gyrokinetic codes for populating the database. Neural network topology complexity favourably scales linearly with input dimensionality. However, we estimate that uniform density population of the input dimensions, as carried out in this work, is feasible up to N$\sim$10. This is due to constraints on the NN training and quasilinear database calculation times. For higher dimensionalities, a training set which captures the natural correlations of the input parameters is then vital. This can be done by basing the training sets on experimental parameters and reasonable extrapolations thereof. This is a feasible goal, as also evidenced in Ref.~\cite{mene14}, and this work is ongoing. 

\textit{Acknowledgments}.--
This work is part of the research programme `Fellowships for Young Energy Scientists' (YES!) of the Foundation for Fundamental Research on Matter (FOM), which is financially supported by the Netherlands Organisation for Scientific Research (NWO). The authors greatly thank O. Meneghini, S. Smith, U. von Toussaint and P. Xanthopolous for inspiring discussions.

\section*{References}

\bibliographystyle{unsrt}
\bibliography{neuralbib}

\end{document}